\documentclass[10pt,a4paper]{article}
\usepackage{amssymb,amsmath,amscd,amsthm}
\usepackage{bbm}
\usepackage{graphicx}
\usepackage{braket}
\usepackage[usenames]{color}

\usepackage{wrapfig}
\usepackage{framed}
\usepackage{ascmac}
\usepackage{color}
\usepackage{dcolumn}
\usepackage{amsfonts}
\usepackage{bbm} %% for bb 1
\newcommand{\QED}{$\blacksquare$}

\newtheorem{theorem}{Theorem}
\newtheorem{lmm}[theorem]{Lemma}

\newtheorem{pro}[theorem]{Proposition}
\newtheorem{df}[theorem]{Definition}
\newtheorem{rmk}[theorem]{Remark}

\newcommand\calA{{\cal A}}

\newcommand\calD{{\cal D}}

\newcommand\calP{{\cal P}}

\newcommand\calW{{\cal W}}

\newcommand\frah{{\frak h}}
\newcommand\fraH{{\frak H}}
\newcommand\fraK{{\frak K}}

\newcommand\dprime{{\prime\prime}}

\newcommand\bbR{{\mathbb R}}
\newcommand\bbN{{\mathbb N}}
\newcommand\bbZ{{\mathbb Z}}
\newcommand\bbC{{\mathbb C}}

\newcommand\0{\{ 0 \}}

\newcommand{\abs}[1]{{\left|#1\right|}}
\newcommand{\norm}[1]{{\left\Vert#1\right\Vert}}

\newcommand{\rbk}[1]{{\left(#1\right)}}
\newcommand{\sbk}[1]{{\left[#1\right]}}
\newcommand{\cbk}[1]{{\left\{#1\right\}}}

\newcommand{\innpro}[1]{{\left\langle#1\right\rangle}}

\begin{document}

\newpage\thispagestyle{empty}
\begin{center}
{\huge\bf
Remarks on BEC on Graphs}
\\
\bigskip\bigskip
\bigskip\bigskip
{\Large Tomohiro Kanda}
\\
Graduate School of Mathematics, Kyushu University,
\\
744 Motoka, Nishi-ku, Fukuoka 819-0395, JAPAN
\\
 t-kanda@math.kyushu-u.ac.jp
\\
\end{center}
\bigskip\bigskip\bigskip\bigskip
\bigskip\bigskip\bigskip\bigskip
{\bf Abstract:}
We consider Bose--Einstein condensation (BEC) on graphs with transient adjacency matrix.
We prove the equivalence of BEC and non-factoriality of a quasi-free state.
Moreover, quasi-free states exhibiting BEC decompose into generalized coherent states. 
We review necessary and sufficient conditions that a quasi-free state is faithful, factor, and pure and quasi-free states are quasi-equivalent, including the paper of H. Araki and M. Shiraishi (1971/72), H. Araki (1971/72), and H. Araki and S. Yamagami (1982).
Using their formats and results, we prove necessary and sufficient conditions that a generalized coherent state is faithful, factor, and pure and generalized coherent states are quasi-equivalent as well.

%This is also for double spacing
\newpage

\setcounter{theorem}{0}
\setcounter{equation}{0}
\section{Introduction} \label{sec:introduction}
In \cite{Matsui}, T. Matsui studied the condition for Bose--Einstein condensation (BEC for short.) in terms of the random walk on a graph.
In \cite{Fidaleo11}, F. Fidaleo, D. Guido, and T. Isola, in \cite{Fidaleo12} and \cite{Fidaleo15}, F. Fidaleo studied some spectral properties of the adjacency matrix of graphs and BEC. 
They obtained the criterion for BEC on graphs.
In \cite{LewisPule}, J. T. Lewis and J. V. Pul\`{e} obtained the non-factoriality of a quasi-free state exhibiting BEC in $L^2(\bbR^3)$ case.
However, in case of graphs, BEC implies non-factoriality of a quasi-free states exhibiting BEC is not clear, thus, we study a quasi-free state exhibiting BEC and prove the equivalence of the occurrence of BEC and non-factoriality of a quasi-free state.
Moreover, we give factor decomposition of quasi-free states exhibiting BEC into generalized coherent states which are factor and mutually disjoint (Theorem \ref{theorem:factor decomp h}.).
Generalized coherent states are generalization of coherent states in the following sense.
Let $\frah$ be a Hilbert space and $\sigma$ be the symplectic form defined by $\sigma(f,g) := {\rm Im}\innpro{f,g}_{\frah}$.
In mathematics, a coherent state $\varphi$ on the Weyl CCR algebra $\calW(\frah, \sigma)$ is given by
\begin{equation}
\varphi(W(f)) = \exp\{ - \norm{f}^2_{\frah}/4 + i{\rm Re}\lambda(f) \}
\end{equation}
for each $f \in \frah$, where $W(f)$, $f \in \frah$, are unitaries which generate $\calW(\frah, \sigma)$ and $\lambda$ is a $\bbC$-linear functional on $\frah$.
(See \cite[Theorem 3.1.]{HoneggerArmin}.)
A state $\varphi$ on $\calW(\frah, \sigma)$ is a generalized coherent state, if there exists a positive semi-definite sesquilinear form $S$ on $\frah \times \frah$ and an $\bbR$-linear functional $\lambda : \frah \to \bbR$ such that 
\begin{equation}
\varphi(W(f)) = \exp\{ -S(f,f)/4 + i \lambda(f) \}, \quad f \in \frah.
\end{equation}

In Section \ref{sec:Quasi-free states}, we review works of H. Araki and M. Shiraishi \cite{Araki1}, H. Araki \cite{Araki2}, and H. Araki and S. Yamagami \cite{ArakiYamagami}.
In \cite{Araki1}, H. Araki and M. Shiraishi and in \cite{Araki2}, H. Araki considered quasi-free states on the CCR algebra and obtained a condition that a quasi-free state is  faithful, factor, and pure.
In \cite{ArakiYamagami}, H. Araki and S. Yamagami got necessary and sufficient conditions that quasi-free states are quasi-equivalent.
In \cite{ManuceauVerbeure68}, J. Manuceau and A. Verbeure and in \cite{ManuceauRoccaTestard69}, J. Manuceau, F. Rocca, and D. Testard obtained a condition that  a quasi-free state on the Weyl CCR algebra is pure and factor. In \cite{vanDaele}, A. van Daele obtained conditions of quasi-equivalence of quasi-free states on the Weyl CCR algebra as well.
To consider conditions of factoriality, purity and faithfulness of a generalized coherent state and conditions of quasi-equivalence of generalized coherent states in a unified framework, we use formats in \cite{Araki1}, \cite{Araki2}, and \cite{ArakiYamagami}.

In Section \ref{sec:coherent states}, we consider generalized coherent states on the Weyl CCR algebra.
We prove necessary and sufficient conditions that a generalized coherent state is faithful, factor, and pure and necessary and sufficient conditions that generalized coherent states are quasi-equivalent.
Moreover, we give an explicit form of factor decomposition of non-factor generalized coherent state.
In \cite{Honegger90}, R. Honegger considered the decomposition of gauge-invariant quasi-free states.
In the present paper, we only assume that a state on the Weyl CCR algebra is quasi-free or generalized coherent.

In Section \ref{sec:graphs}, we review works of F. Fidaleo \cite{Fidaleo15} and consider the non-factoriality of quasi-free states with BEC.
We show that a quasi-free state exhibiting BEC is non-factor and such state decomposes into generalized coherent states which are mutually disjoint.
In \cite{PuleVerbeureZagrebnov}, J. V. Pul\'{e}, A. F. Verbeure, and V. A. Zagrebnov considered inhomogeneous BEC on $L^2(\bbR^\nu)$, $\nu \geq 1$, and obtained that the occurrence of BEC implies spontaneous symmetry breaking and an equilibrium state exhibiting BEC decompose into periodic states.
In \cite{Matsui}, T. Matsui obtained that the occurrence of BEC implies spontaneous symmetry breaking in case of graphs with some assumptions (See \cite[Assumption 1.1.]{Matsui}.) as well.
In the present paper, generalized coherent states appeared in factor decomposition of a quasi-free state are not periodic.
Thus, we give another decomposition of a quasi-free state exhibiting BEC.

\setcounter{theorem}{0}
\setcounter{equation}{0}
\section{Preliminaries} \label{sec:Quasi-free states}
In this section, we review works of H. Araki and M. Shiraishi \cite{Araki1}, H. Araki \cite{Araki2}, and H. Araki and S. Yamagami \cite{ArakiYamagami}.
In \cite{Araki1}, H. Araki and M. Shiraishi and in \cite{Araki2}, H. Araki considered quasi-free states on the CCR algebra and obtained necessary and sufficient conditions that a quasi-free state is factor, pure, and faithful.
In \cite{ArakiYamagami}, H. Araki and S. Yamagami obtained necessary and sufficient conditions that quasi-free states are quasi-equivalent.
We use facts presented in this section to consider necessary and sufficient conditions that a generalized coherent state is factor, pure, and faithful and generalized coherent states are quasi-equivalent and to prove non-factoriality of quasi-free states exhibiting BEC.

\subsection{Some Properties of a Quasi-free state} \label{sec:sub Quasi-free states}
Let $\tilde{K}$ be a $\bbC$-linear space and $\gamma_{\tilde{K}} : \tilde{K} \times \tilde{K} \to \bbC$ be a sesquilinear form.
Let $\Gamma_{\tilde{K}}$ be an anti-linear involution ($\Gamma_{\tilde{K}}^2=\mathbbm{1}$) satisfying $\gamma_{\tilde{K}}(\Gamma_{\tilde{K}} f, \Gamma_{\tilde{K}} g) = - \gamma_{\tilde{K}}(g,f)$.
A CCR algebra $\calA(\tilde{K}, \gamma_{\tilde{K}}, \Gamma_{\tilde{K}})$ over $(\tilde{K}, \gamma_{\tilde{K}}, \Gamma_{\tilde{K}})$ is the quotient of the complex $*$-algebra generated by $B(f)$, $f \in \tilde{K}$, its adjoint $B(f)^*$, $f \in \tilde{K}$ and an identity over the following relations:
\begin{enumerate}
\item $B(f)$ is complex linear in $f$,
\item $B(f)^* B(g) - B(g) B(f)^* = \gamma_{\tilde{K}}(f,g) \mathbbm{1}$, 
\item $B(\Gamma_{\tilde{K}} f)^* = B(f)$.
\end{enumerate}

Any linear operator $P$ on $\tilde{K}$ satisfying
\begin{enumerate}
\item $P^2 = P$,
\item $\gamma_{\tilde{K}}(Pf,g) > 0$, if $Pf \neq 0$,
\item $\gamma_{\tilde{K}}(P f, g) = \gamma_{\tilde{K}}(f, Pg)$,
\item $\Gamma_{\tilde{K}}P \Gamma_{\tilde{K}} = 1 - P$,
\end{enumerate}
is called a basis projection.

Let $\frah$ be a complex pre-Hilbert space.
A CCR ($*$-)algebra $\calA_{{\rm CCR}}(\frah)$ over $\frah$ is the quotient of the $*$-algebra generated by $a^\dagger(f)$ and $a(f)$, $f \in \frah$, and an identity by the following relations:
\begin{enumerate}
\item $a^\dagger(f)$ is complex linear in $f$,
\item $(a^\dagger(f))^* = a(f)$, 
\item $\sbk{ a(f), a^\dagger (g) } = (f,g)_{\frah} \mathbbm{1}$ and $\sbk{ a^\dagger(f), a^\dagger(g) } = 0 = \sbk{a(f), a(g)}$.
\end{enumerate}

Let $P$ be a basis projection.
Then the mapping $\alpha(P)$ from $\calA(\tilde{K}, \gamma_{\tilde{K}}, \Gamma_{\tilde{K}})$ to $\calA_{{\rm CCR}}(P\tilde{K})$ defined by 
\begin{eqnarray}
\alpha(P)(B(f_1) B(f_2) \cdots B(f_n)) &=& (\alpha(P) B(f_1)) (\alpha(P) B(f_2)) \cdots (\alpha(P) B(f_n)) \nonumber \\
\alpha(P)B(f) &=& a^\dagger(Pf) + a(P\Gamma_{\tilde{K}} f) \label{eq:relationships between B and a} 
\end{eqnarray}
is a $*$-isomorphism of $\calA(\tilde{K}, \gamma_{\tilde{K}}, \Gamma_{\tilde{K}})$ onto $\calA_{{\rm CCR}}(P\tilde{K})$.

Let $\calA$ be a $*$-algebra with identity.
A linear functional $\varphi$ on $\calA$ is said to be state, if $\varphi$ satisfies $\varphi(A^* A) \geq 0$, $A \in \calA$, and $\varphi(\mathbbm{1}) = 1$.
For a state $\varphi$ on $\calA$, we have the GNS-representation space $(\fraH_\varphi, \pi_\varphi, \xi_\varphi)$ associated with $\varphi$.
We set ${\rm Re} \tilde{K} := \set{f \in \tilde{K} | \Gamma_{\tilde{K}} f = f}$.
Then $f \in {\rm Re} \tilde{K}$ if and only if $B(f)^* = B(f)$.

On ${\rm Re} \tilde{K}$, the operators $B(f)$, $f \in {\rm Re} \tilde{K}$, correspond to field operators.
Moreover, $a^\dagger(f)$ and $a(f)$ correspond to the creation operators and the annihilation operators.
We give examples of $\tilde{K}$, $\gamma_{\tilde{K}}$, and $\Gamma_{\tilde{K}}$ in Section \ref{sec:coherent states} and \ref{sec:graphs}.

Let $\varphi$ be a state on $\calA(\tilde{K}, \gamma_{\tilde{K}}, \Gamma_{\tilde{K}})$ such that $\pi_\varphi(B(f))$ is essentially self-adjoint for all $f \in {\rm Re} \tilde{K}$.
Then we put $W_\varphi(f) = \exp(i \pi_{\varphi}(B(f)))$, $f \in {\rm Re} \tilde{K}$.
Such state $\varphi$ is said to be regular if $W_\varphi(f)$ satisfies the Weyl--Segal relations:
\begin{equation}
W\varphi(f) W_\varphi(g) = \exp(-\gamma_{\tilde{K}}(f,g)/2) W_\varphi(f + g), \quad f,g \in {\rm Re}\tilde{K}. \label{eq:WeylSegal}
\end{equation}
In general, the Weyl CCR algebra is the universal ${\rm C}^*$-algebra generated by unitaries $W(f)$, $f \in {\rm Re} \tilde{K}$, which satisfy (\ref{eq:WeylSegal}) and we denote $\calW({\rm Re}\tilde{K}, \gamma_{\tilde{K}})$ the Weyl CCR algebra. (See also \cite[Theorem 5.2.8.]{BratteliRobinsonII}.)

A state $\varphi$ on $\calA(\tilde{K}, \gamma_{\tilde{K}}, \Gamma_{\tilde{K}})$ is said to be quasi-free, if $\varphi$ satisfies the following equations:
\begin{eqnarray}
& & \varphi(B(f_1) \cdots B(f_{2n -1})) = 0, \nonumber\\
& & \varphi(B(f_1) \cdots B(f_{2n})) = \sum \prod_{j = 1}^n \varphi(B(f_{s(j)}) B(f_{s(j + n)})), \label{eq:def of quasi-free on star-alg}
\end{eqnarray}
where $n \in \bbN$ and the sum is over all permutations $s$ satisfying $s(1) < s(2) < \cdots < s(n), s(j) < s(j + n)$, $j = 1,2, \cdots, n$.
For any quasi-free state $\varphi$ over $\calA(\tilde{K}, \gamma_{\tilde{K}}, \Gamma_{\tilde{K}})$, the sesquilinear form $S_{\tilde{K}} : \tilde{K} \times \tilde{K} \to \bbC$ defined by
\begin{equation}
S_{\tilde{K}}(f,g) = \varphi(B(f)^*B(g)), \quad f,g \in \tilde{K} \label{eq:two-pt func def}
\end{equation}
is positive semi-definite and satisfies 
\begin{equation}
\gamma_{\tilde{K}}(f,g) = S_{\tilde{K}}(f,g) - S_{\tilde{K}}(\Gamma g, \Gamma f), \quad f,g \in\tilde{K}. \label{eq:satisfying eq of S}
\end{equation}
(See \cite[Lemma 3.2.]{Araki1}.)
Any quasi-free state on $\calA(\tilde{K}, \gamma_{\tilde{K}}, \Gamma_{\tilde{K}})$ determines the positive semi-definite sesquilinear form $S$, which satisfies the equation (\ref{eq:satisfying eq of S}).
Conversely, for any positive semi-definite sesquilinear form $S_{\tilde{K}}$ on $\tilde{K} \times \tilde{K}$ satisfying (\ref{eq:satisfying eq of S}), there exists a unique quasi-free state $\varphi$ satisfying (\ref{eq:two-pt func def}) and $\varphi$ is regular.
(See \cite[Lemma 3.5.]{Araki1}.)
Thus, there exists a one-to-one correspondence between a positive semi-definite sesquilinear form $S_{\tilde{K}}$ on $\tilde{K} \times \tilde{K}$ and a quasi-free state $\varphi$ on $\calA(\tilde{K}, \gamma_{\tilde{K}}, \Gamma_{\tilde{K}})$.
We denote the quasi-free state on $\calA(\tilde{K}, \gamma_{\tilde{K}}, \Gamma_{\tilde{K}})$ determined by a positive semi-definite sesquilinear form $S_{\tilde{K}}$ by $\varphi_S$ defined in (\ref{eq:two-pt func def}).
We define the positive semi-definite form $(\cdot,\cdot)_S$ on $\tilde{K} \times \tilde{K}$ by the following equation:
\begin{equation}
(f,g)_S := S_{\tilde{K}}(f,g) + S_{\tilde{K}}(\Gamma_{\tilde{K}} g, \Gamma_{\tilde{K}} f), \quad f,g \in \tilde{K}. \label{eq:def of inn pro induced by S}
\end{equation}
We set $N_S := \set{ f \in \tilde{K} | \norm{f}_S = 0}$, where $\norm{f}_S = (f,f)_S^{1/2}$.
We denote the completion of $\tilde{K} / N_S$ with respect to the norm $\norm{\cdot}_S$ by $K$.
Since $S_{\tilde{K}}(f, f) \leq \norm{f}^2_S$, $\abs{\gamma_{\tilde{K}}(f,f)} \leq \norm{f}^2_S$, and $\norm{\Gamma_{\tilde{K}} f}_S = \norm{f}_S$ for any $f \in \tilde{K}$, we can extend the sesquilinear form $S_{\tilde{K}}$ and $\gamma_{{\tilde{K}}}$ to the sesquilinear form on $K \times K$ and the operator $\Gamma_{\tilde{K}}$ to the operator on $K$.
We denote the extensions of $S_{\tilde{K}}$, $\gamma_{\tilde{K}}$, and $\Gamma_{\tilde{K}}$ by $S_K$, $\gamma_K$, and $\Gamma_K$, respectively.
We define the bounded operators $S_K$ and $\gamma_K$ on $K$ by the following equations:
\begin{eqnarray}
(\xi, S_K \eta)_S &=& S_K(\xi,\eta), \label{eq:def of S}\\
(\xi, \gamma_K \eta)_S &=& \gamma_K(\xi,\eta), \quad \xi, \eta \in K. \label{eq:def of gamma}
\end{eqnarray}
A quasi-free state $\varphi_S$ is said to be Fock type if $N_S = \{ 0 \}$ and the spectrum of the operator $S_K$ defined in (\ref{eq:def of S}) is contained in $\{ 0, 1/2, 1\}$. 
For any positive semi-definite sesquilinear form $S_{\tilde{K}}$ on $\tilde{K} \times \tilde{K}$, we can construct a Fock type state as follows.
Let $\tilde{L} = K \oplus K$.
For $\xi_1, \xi_2, \eta_1, \eta_2 \in K$, we set
\begin{eqnarray}
\gamma_L(\xi_1 \oplus \xi_2, \eta_1 \oplus \eta_2) &=& (\xi_1, \gamma_K \eta_1)_S - (\xi_2, \gamma_K \eta_2)_S, \label{eq:gammaL} \\
\widetilde{\Gamma_L} &=& \Gamma_K \oplus \Gamma_K, \label{eq:GammaL} \\
(\xi_1 \oplus \xi_2, \eta_1 \oplus \eta_2)_{L} &=& (\xi_1, \eta_1)_S + (\xi_2, \eta_2)_S + 2( \xi_1, S_K^{1/2}(\mathbbm{1} - S_K)^{1/2} \eta_2 )_S \nonumber\\
& & + 2 ( \xi_2, S_K^{1/2}(\mathbbm{1} - S_K)^{1/2} \eta_1 )_S.  \label{eq:innproL}
\end{eqnarray}
Let $N_L = \set{\xi \in \tilde{L} | (\xi,\xi)_L = 0}$.
Then we denote the completion of $\tilde{L} / N_L$ with respect to the norm $\norm{ \cdot }_L$ by $L$.
We define the bounded operators $\gamma_{L}$ and $\Pi_L$ on $L$ satisfying
\begin{eqnarray}
(\xi, \gamma_L \eta)_L &=& \gamma_L(\xi, \eta), \quad \xi, \eta \in L. \label{eq:opgammaL} \\
\Pi_L &=& \frac{1}{2}(\mathbbm{1} + \gamma_L). \label{eq:PiL}
\end{eqnarray}
Then the spectrum of $\Pi_L$ on $L$ is contained in $\{ 0, 1/2, 1 \}$. 
(See \cite[Lemma 5.8.]{Araki1} and \cite[Lemma 6.1.]{Araki1}.)
Moreover the following three lemmas hold:
\begin{lmm} {\rm \cite[Corollary 6.2.]{Araki1}} \label{lmm:araki1}
The map $f \in \tilde{K} \mapsto [f] \in L$, where $[f] : = (f \oplus 0) + N_L$, induces a $*$-homomorphism $\alpha_{\tilde{K}}$ of $\calA(\tilde{K}, \gamma_{\tilde{K}}, \Gamma_{\tilde{K}})$ into $\calA(L, \gamma_L, \Gamma_L)$.
The restriction of a Fock type state $\varphi_{\Pi_L}$ of $\calA(L, \gamma_L, \Gamma_L)$ to $\alpha_{\tilde{K}}(\calA(\tilde{K}, \gamma_{\tilde{K}}, \Gamma_{\tilde{K}}))$ gives a quasi-free state $\varphi_S$ of $\calA(\tilde{K}, \gamma_{\tilde{K}}, \Gamma_{\tilde{K}})$ through $\varphi_{\Pi_L}(\alpha_{\tilde{K}}(A)) = \varphi_S(A)$.
\end{lmm}

\begin{lmm} {\rm \cite[Lemma 2.3.]{Araki2}} \label{lmm:araki2}
Let $R_S$ be the von Neumann algebra generated by spectral projections of all $\pi_{\Pi_L}(B(f))$, $f \in {\rm Re}{\tilde{K}}$, on the GNS representation space $(\fraH_{\Pi_L}, \pi_{\Pi_L}, \xi_{\Pi_L})$ of $\calA(L, \gamma_L, \Gamma_L)$ associated with $\varphi_{\Pi_L}$.
Then the following conditions are equivalent:
\begin{enumerate}
\item The GNS cyclic vector $\xi_{\Pi_L}$ is cyclic for $R_S$.
\item The GNS cyclic vector $\xi_{\Pi_L}$ is separating for $R_S$.
\item The operator $S_K$ on $K$ does not have an eigenvalue $0$.
\item The operator $S_K$ on $K$ does not have an eigenvalue $1$.
\end{enumerate}
\end{lmm}

\begin{lmm} {\rm \cite[Lemma 2.4.]{Araki2}} \label{lmm:araki3}
The center of $R_S$ is generated by $\exp(i \pi_{\Pi_L}(B(h)))$, $h \in {\rm Re} \overline{(E_0 K \oplus 0)}^L$, where $E_0$ is the spectral projection of $S_K$ for $ 1/2 $ and $\overline{(E_0 K \oplus 0)}^L$ is the closure of $E_0 K \oplus 0$ with respect to the norm $\norm{\cdot}_{L}$.
In particular, $R_S$ is factor if and only if $K_0 = E_0 K = \0$.
\end{lmm}

\subsection{Quasi-equivalence of Quasi-free states}  \label{sec:Quasi-equivalence of Quasi-free states}
We recall the definitions of quasi-equivalence of representations and states.
\begin{df} {\rm \cite[Definition 6.1.]{Araki2}} \label{df:quasi-equivalence}
Let $\pi_{S_1}$ and $\pi_{S_2}$ be representations associated with quasi-free states $\varphi_{S_1}$ and $\varphi_{S_2}$ on $\calA(\tilde{K}, \gamma_{\tilde{K}}, \Gamma_{\tilde{K}})$, respectively.
The representations $\pi_{S_1}$ and $\pi_{S_2}$ are said to be quasi-equivalent, if there exists an isomorphism $\tau$ from $R_{S_1} = \set{W_{S_1}(f) | f \in {\rm Re} \tilde{K}}^\dprime$ onto $R_{S_2} = \set{W_{S_2}(f) | f \in {\rm Re} \tilde{K}}^\dprime$ such that
\begin{equation}
\tau(W_{S_1}(f)) = W_{S_2}(f), \quad f \in {\rm Re} \tilde{K}, \label{eq:quasi-eq iso}
\end{equation}
where $W_{S_1}(f) = \exp(i \pi_{S_1}(B(f)))$ and $W_{S_2}(f) = \exp(i \pi_{S_2}(B(f)))$.
Let $\varphi_{S_1}$ and $\varphi_{S_2}$ be quasi-free states on $\calA(\tilde{K}, \gamma_{\tilde{K}}, \Gamma_{\tilde{K}})$.
The states $\varphi_{S_1}$ and $\varphi_{S_2}$ are said to be quasi-equivalent, if for each GNS-representations $(\fraH_{S_i}, \pi_{S_i})$, $i = 1,2$ associated with $\varphi_{S_i}$, respectively, are quasi-equivalent.
\end{df}
This definition is equivalent to the definition of quasi-equivalence of states on a ${\rm C}^*$-algebra.
(See \cite[Definition 2.4.25.]{BratteliRobinsonI} and \cite[Theorem 2.4.26.]{BratteliRobinsonI}.)

Let $\varphi_{S_1}$ and $\varphi_{S_2}$ be quasi-free states on $\calA(\tilde{K}, \gamma_{\tilde{K}}, \Gamma_{\tilde{K}})$.
In \cite{ArakiYamagami}, H. Araki and S. Yamagami showed the following theorem:
\begin{theorem} {\rm \cite[Theorem]{ArakiYamagami}} \label{theorem:ArakiYamagami}
Two quasi-free states $\varphi_{S_1}$ and $\varphi_{S_2}$ on $\calA(\tilde{K}, \gamma_{\tilde{K}}, \Gamma_{\tilde{K}})$ are quasi-equivalent if and only if the following conditions hold:
\begin{enumerate}
\item The topologies induced by $\norm{\cdot}_{S_1}$ and $\norm{\cdot}_{S_2}$ are equal.
\item Let $K$ be the completion of $\tilde{K}$ with respect to the topology $\norm{\cdot}_{S_1}$ or $\norm{\cdot}_{S_2}$.
Then $S_1^{1/2} - S_2^{1/2}$ is in the Hilbert--Schmidt class on $K$, where the $S_1$ and $S_2$ are operators on $K$ defined in {\rm (\ref{eq:def of S})}.
\end{enumerate}
\end{theorem}

%%%%%%%%%%%%%%%%%%%%%%%%%%%%%%%%%%%%%%%%%%%%%%%%%%
%%%%%%%%%%%%%%%%%%%%%%%%%%%%%%%%%%%%%%%%%%%%%%%%%%
%%%%%%%%%%%%%%%%%%%%%%%%%%%%%%%%%%%%%%%%%%%%%%%%%%
%%%%%%%%%%%%%%%%%%%%%%%%%%%%%%%%%%%%%%%%%%%%%%%%%%
%%%%%%%%%%%%%%%%%%%%%%%%%%%%%%%%%%%%%%%%%%%%%%%%%%
%%%%%%%%%%%%%%%%%%%%%%%%%%%%%%%%%%%%%%%%%%%%%%%%%%
%%%%%%%%%%%%%%%%%%%%%%%%%%%%%%%%%%%%%%%%%%%%%%%%%%
%%%%%%%%%%%%%%%%%%%%%%%%%%%%%%%%%%%%%%%%%%%%%%%%%%
%%%%%%%%%%%%%%%%%%%%%%%%%%%%%%%%%%%%%%%%%%%%%%%%%%
%%%%%%%%%%%%%%%%%%%%%%%%%%%%%%%%%%%%%%%%%%%%%%%%%%
%%%%%%%%%%%%%%%%%%%%%%%%%%%%%%%%%%%%%%%%%%%%%%%%%%
%%%%%%%%%%%%%%%%%%%%%%%%%%%%%%%%%%%%%%%%%%%%%%%%%%
%%%%%%%%%%%%%%%%%%%%%%%%%%%%%%%%%%%%%%%%%%%%%%%%%%
%%%%%%%%%%%%%%%%%%%%%%%%%%%%%%%%%%%%%%%%%%%%%%%%%%
%%%%%%%%%%%%%%%%%%%%%%%%%%%%%%%%%%%%%%%%%%%%%%%%%%
%%%%%%%%%%%%%%%%%%%%%%%%%%%%%%%%%%%%%%%%%%%%%%%%%%
%%%%%%%%%%%%%%%%%%%%%%%%%%%%%%%%%%%%%%%%%%%%%%%%%%
%%%%%%%%%%%%%%%%%%%%%%%%%%%%%%%%%%%%%%%%%%%%%%%%%%
%%%%%%%%%%%%%%%%%%%%%%%%%%%%%%%%%%%%%%%%%%%%%%%%%%
%%%%%%%%%%%%%%%%%%%%%%%%%%%%%%%%%%%%%%%%%%%%%%%%%%
%%%%%%%%%%%%%%%%%%%%%%%%%%%%%%%%%%%%%%%%%%%%%%%%%%
%%%%%%%%%%%%%%%%%%%%%%%%%%%%%%%%%%%%%%%%%%%%%%%%%%
%%%%%%%%%%%%%%%%%%%%%%%%%%%%%%%%%%%%%%%%%%%%%%%%%%
%%%%%%%%%%%%%%%%%%%%%%%%%%%%%%%%%%%%%%%%%%%%%%%%%%
\setcounter{theorem}{0}
\setcounter{equation}{0}
\section{Generalized Coherent states} \label{sec:coherent states}
In this section, we consider generalized coherent states on the Weyl CCR algebra.
Using facts in the previous section, we give necessary and sufficient conditions that a generalized coherent state is factor, pure, and faithful and generalized coherent states are quasi-equivalent as well.

\subsection{The Weyl CCR algebra} \label{sec:Weyl CCR algebra}
Let $V$ be an $\bbR$-linear space with a symplectic form $\sigma : V \times V \to \bbR$, i.e., $\sigma$ is a bilinear form on $V$ and satisfy the following relations:
\begin{equation}
\sigma(f,g) = - \sigma(g,f), \quad f, g \in V. \label{eq:symplectic relation}
\end{equation}
We assume that there exists an operator $J$ on $V$ with the properties
\begin{equation}
\sigma(Jf, g) = - \sigma(f, Jg), \quad J^2 = -1, \quad \label{eq:complexification}
\end{equation}
then $V$ is a $\bbC$-linear space with scalar multiplication defined by
\begin{equation}
(c_1 + i c_2) f = c_1 f + c_2 Jf, \quad c_1, c_2 \in \bbR, \, f \in V. \label{eq:complexification of V}
\end{equation}
Then we define the complexification $V^\bbC$ of $V$ by (\ref{eq:complexification of V}).
We set $(f + ig)^* = f - ig$ for $f,g \in V$.
We fix a symplectic space $(V, \sigma)$ with an operator $J$ satisfying (\ref{eq:complexification}).
We puts $\tilde{K} = V^\bbC$,
\begin{eqnarray}
\Gamma_{\tilde{K}} f &=& f^*, \quad f \in \tilde{K}, \nonumber\\
\gamma_{\tilde{K}}(f,g) &=& \frac{1}{2} \cbk{ \sigma(f, Jg) + i \sigma(f,g) - \sigma(g^*, Jf^*) - i \sigma(g^*,f^*)}, \quad f,g \in \tilde{K}. \label{eq:case of symplectic space}
\end{eqnarray}
Then on the GNS-representation space $(\fraH_\varphi, \pi_\varphi)$ associated with a regular state $\varphi$ on $\calA(\tilde{K}, \gamma_{\tilde{K}}, \Gamma_{\tilde{K}})$, $\calW({\rm Re}\tilde{K}, \gamma_{\tilde{K}}) = \calW(V, \sigma)$. 
Moreover, $\pi_\varphi(B(f))$, $f \in {\rm Re}\tilde{K}$, correspond to filed operators.
We define the annihilation operators $a(f)$ and the creation operators $a^\dagger(f)$ on $\fraH_\varphi$ by the following equation:
\begin{equation}
a_\varphi(f):= \{ \pi_\varphi(B(f)) + i \pi_\varphi(B(if)) \}/\sqrt{2}, \quad a^\dagger(f) := \{ \pi_\varphi(B(f)) - i \pi_\varphi(B(if)) \}/\sqrt{2},\label{eq:annihilation and creation}
\end{equation}
for any  $f \in {\rm Re} \tilde{K}$.

In this section, we identify the Weyl CCR algebra $\calW(V, \sigma)$ with a regular state $\varphi$ and $\calA(\tilde{K}, \gamma_{\tilde{K}}, \Gamma_{\tilde{K}})$ with $\varphi$, where $\tilde{K}$, $\gamma_{\tilde{K}}$ and $\Gamma_{\tilde{K}}$ defined in (\ref{eq:case of symplectic space}).

\subsection{Generalized coherent states}
For an $\bbR$-linear functional $\lambda: V \to \bbR$, there exists a $*$-automorphism $\tau_\lambda$ on $\calW(V, \sigma)$ defined by
\begin{equation}
\tau_\lambda(W(f)) := e^{i\lambda(f)} W(f), \quad f \in V. \label{eq:phase transition}
\end{equation}
Let $\varphi_S$ be a quasi-free state on $\calW(V, \sigma)$.
Then we define the generalized coherent state $\varphi_{S, \lambda}$ by the following equation:
\begin{equation}
\varphi_{S, \lambda}(W(f)) := \varphi_S \circ \tau_\lambda(W(f)) = e^{i \lambda(f)} \varphi_S(W(f)), \quad f \in V. \label{eq:def of coherent state}
\end{equation}
We sets $N_S = \set{ f \in V^\bbC | \norm{f}_S = 0}$, where $\norm{\cdot}_S = (\cdot,\cdot)^{1/2}_S$ is the semi-norm defined in (\ref{eq:def of inn pro induced by S}) and $V^\bbC_S$ is the completion of $V^\bbC/N_S$ by the norm $\norm{\cdot}_S$.
We denote the GNS-representation space with respect to $\varphi_S$ and $\varphi_{S, \lambda}$ by $(\fraH_S, \pi_S, \xi_S)$ and $(\fraH_{S, \lambda}, \pi_{S,\lambda}, \xi_{S, \lambda})$, respectively.

\begin{lmm} \label{lmm:equality of von Neumann algebras}
Let $\varphi_S$ and $\varphi_{S, \lambda}$ be a quasi-free state and a generalized coherent state on $\calW(V, \sigma)$, respectively.
Then 
\begin{equation}
R_S = R_{S, \lambda}, \label{eq:von Neuamnn algebras w.r.t. q-f and coh state}
\end{equation}
where $R_S$ and $R_{S, \lambda}$ is the von Neumann algebra generated by $\set{ \pi_S(W(f)) | f \in V }$ and $\set{ \pi_{S, \lambda} (W(f)) | f \in V }$, respectively.
\end{lmm}

\noindent {\bf Proof.}
Since $\varphi_S$ is regular, there exist self-adjoint operators $\Psi_S(f)$, $f \in V$ such that $\pi_S(W(f)) = \exp(i \Psi_S(f))$.
By definition of generalized coherent states, we have $\pi_{S, \lambda}(W(f)) = e^{i\lambda(f)}\pi_{S}(W(f))$ and $(\fraH_{S, \lambda}, \pi_{S, \lambda}, \xi_{S, \lambda}) = (\fraH_S, \pi_{S, \lambda}, \xi_{S})$.
On $\fraH_S$, we have 
\begin{eqnarray}
& & \set{\pi_S(W(f)) | f \in V}^\dprime = \set{ e^{i \lambda(f)} \pi_S(W(f)) | f \in V}^\dprime \nonumber\\
&=& \set{\pi_{S, \lambda}(W(f)) | f \in V}^\dprime. \label{eq:equality of von Neumann algebra}
\end{eqnarray}
Thus, $R_S = R_{S, \lambda}$ by the double commutant theorem.
\QED

\begin{theorem}
Let $\varphi_{S, \lambda}$ be a generalized coherent state on $\calW(V, \sigma)$.
Then $\varphi_{S, \lambda}$ is faithful if and only if $S$ does not have an eigenvalue $0$ on $V^\bbC_S$.
\end{theorem}

\noindent {\bf Proof.}
Note that $\varphi_{S}$ and $\varphi_{S, \lambda}$ has the same GNS cyclic vector space $\xi_{\Pi_L}$.
By Lemma \ref{lmm:araki2}, $\varphi_{S, \lambda}$ is faithful if and only if $S$ does not have an eigenvalue $0$ on $V^\bbC_S$. 
\QED

\begin{theorem} \label{theorem:factoriality}
Let $\varphi_{S, \lambda}$ be a generalized coherent state on $\calW(V, \sigma)$.
Then $\varphi_{S, \lambda}$ is factor if and only if $S$ does not have an eigenvalue $1/2$ on $V^\bbC_S$.
\end{theorem}

\noindent {\bf Proof.}
By Lemma \ref{lmm:araki3} and Lemma \ref{lmm:equality of von Neumann algebras}, we have the statement.
\QED

\begin{theorem} 
Let $(V, \sigma)$ be a non-degenerate symplectic space and $\varphi_{S, \lambda}$ be a generalized coherent state on $\calW(V, \sigma)$.
Then $\varphi_{S, \lambda}$ is pure if and only if $S$ is a basis projection.
\end{theorem}

\noindent {\bf Proof.}
If $S$ is a basis projection, then by Lemma \ref{lmm:equality of von Neumann algebras} and \cite[Lemma 5.5.]{Araki1} $\varphi_S$ is pure.

We use the notation in Section \ref{sec:Quasi-free states}.
Thus, $\tilde{K} = V^\bbC$, $K = V^\bbC_S$, and $L$ is the completion of $V^\bbC_S \oplus V^\bbC_S/ N_L$ with respect to the norm $\norm{\cdot}_L$ defined in (\ref{eq:innproL}).
If $\varphi_{S, \lambda}$ is pure, then by Theorem \ref{theorem:factoriality}, $S$ does not have an eigenvalue $1/2$.
Then $\Pi_L$ defined in (\ref{eq:PiL}) does not have an eigenvalue $1/2$ because the eigenspace of $\Pi_L$ with $1/2$ is the completion of the set $\set{ f \oplus f | f \in E_0 K}$ with respect to the norm $\norm{ \cdot }_L$, where $E_0$ is the spectral projection of $S$ onto $\ker(S - 1/2)$.
(See also the proof of (4) of \cite[Lemma 6.1.]{Araki1}.)
Thus, $\Pi_L$ is a basis projection.
Using the notation of \cite[Lemma 5.5.]{Araki1}, we have $R_S = R_{\Pi_L}(H_1)$, with $H_1 = [{\rm Re}\tilde{K}] \oplus 0 \subset L$ and $\overline{H_1} = \overline{{\rm Re} K \oplus 0}^L \oplus 0$.
If $\Pi_L \neq S$, then $K \neq L$.
Thus, we have $R_{\Pi_L}(H_1)^\prime = R_{\Pi_L}(H_1^\perp)$ by \cite[Lemma 5.5.]{Araki1} and $H_1^\perp \neq \0$, where $H_1^\perp$ is the orthogonal complement with respect to the inner product $(\cdot,\cdot)_L$ defined in (\ref{eq:innproL}).
It leads $R_S^\prime \neq \bbC \mathbbm{1}$.
It contradict to the purity of $\varphi_S$.
Thus, $S$ is a basis projection.
\QED

We have necessary and sufficient conditions that a generalized coherent state is faithful, factor, and pure.
Next, we consider the quasi-equivalence of generalized coherent states.

\begin{lmm} \label{lmm:kernel property}
Let $\varphi_{S, \lambda}$ be a generalized coherent state on $\calW(V, \sigma)$.
Then $f \in N_S$ if and only if $\pi_{S, \lambda}(W(f)) = e^{i\lambda(f)} \mathbbm{1}$.
\end{lmm}

\noindent {\bf Proof.}
If $f \in N_S$, then $\varphi_{S}(W(tf)) = 1$ for any $t \in \bbR$.
Thus, by regularity of $\varphi_S$, $\pi_S(W(f)) = \mathbbm{1}$.
By definition of generalized coherent state, $\pi_{S, \lambda}(W(f)) = e^{i \lambda(f)} \mathbbm{1}$.

If $\pi_{S, \lambda}(W(f)) = e^{i\lambda(f)} \mathbbm{1}$, $f \in V$, then $\pi_S(W(f)) = \mathbbm{1}$.
Since $g^* = g$ for any $g \in V$, we have that $(f, f)_S = 0$.
\QED

\begin{lmm} \label{lmm:kernel equality}
Let $\varphi_{S_1, \lambda_1}$ and $\varphi_{S_2, \lambda_2}$ be generalized coherent states on $\calW(V, \sigma)$.
If $\varphi_{S_1, \lambda_1}$ and $\varphi_{S_2, \lambda_2}$ are quasi-equivalent, then $N_{S_1} = N_{S_2}$.
\end{lmm}

\noindent {\bf Proof.}
Since $\varphi_{S_1, \lambda_1}$ and $\varphi_{S_2, \lambda_2}$ are quasi-equivalent, then there exists $\tau : \pi_{S_1, \lambda_1}(\calW(V, \sigma))^\dprime \to \pi_{S_2, \lambda_2}(\calW(V, \sigma))^\dprime$ such that
\begin{equation}
\tau(\pi_{S_1, \lambda_1}(A)) = \pi_{S_2, \lambda_2}(A), \quad A \in \calW(V, \sigma).  \label{eq:qe iso kernel equality}
\end{equation}
If $N_{S_1} \neq N_{S_2}$, then there exists $f \in V^\bbC$ such that $f \in N_{S_1}$ and $f \not \in N_{S_2}$.
Put $h = f+f^*$.
Then $h \in V = {\rm Re}V^\bbC$ and $h \in N_{S_1}$ and $h \not \in N_{S_2}$.
For such $h$, we have
\begin{equation}
\pi_{S_1, \lambda_1} (W(h)) = e^{i \lambda_1(h)} \mathbbm{1} \label{eq:if in kernel kernel equality}
\end{equation}
by Lemma \ref{lmm:kernel property}.
However, we have
\begin{equation}
\pi_{S_2, \lambda_2}(W(h)) = e^{i \lambda_2(h)} \pi_2(W(h)) = \tau(\pi_{S_1, \lambda_1}(W(h))) = e^{i \lambda_1(h)} \mathbbm{1}. \label{eq:contradiction kernel equality}
\end{equation}
It contradict to Lemma \ref{lmm:kernel property}.
\QED

\begin{theorem} \label{theorem:quasi-equivalence of coherent states}
Let $\varphi_{S_1, \lambda_1}$ and $\varphi_{S_2, \lambda_2}$ be generalized coherent states on $\calW(V, \sigma)$.
Then $\varphi_{S_1, \lambda_1}$ and $\varphi_{S_2, \lambda_2}$ are quasi-equivalent if and only if the following conditions hold:
\begin{enumerate}
\item $\norm{\cdot}_{S_1}$ and $\norm{\cdot}_{S_2}$ induce the same topology,
\item $S_1^{1/2} - S_2^{1/2}$ is a Hilbert--Schmidt class operator,
\item $\lambda_1 = \lambda_2$ on $N_{S_1}=N_{S_2}$,
\item $\lambda_1 - \lambda_2$ is continuous with respect to the norm $\norm{\cdot}_{S_1}$ or $\norm{\cdot}_{S_2}$.
\end{enumerate}
\end{theorem}

\noindent {\bf Proof.}
Assume that the topologies induced by $\norm{\cdot}_{S_1}$ and $\norm{\cdot}_{S_2}$ are equivalent, $S_1^{1/2} - S_2^{1/2}$ is Hilbert-Schmidt class, $\lambda_1 - \lambda_2$ is continuous with respect to $\norm{\cdot}_{S_1}$, and $\lambda_1 = \lambda_2$ on $N_{S_1} = N_{S_2}$.
Then $\varphi_{S_1}$ and $\varphi_{S_2}$ are quasi-equivalent by \cite[Theorem]{ArakiYamagami} and $\varphi_{S_1, \lambda_1}$ and $\varphi_{S_2, \lambda_2}$ are quasi-equivalent by continuity of $\lambda_1 - \lambda_2$ and $\lambda_1 = \lambda_2$ on $N_{S_1} = N_{S_2}$.

Next, we assume that $\varphi_{S_1, \lambda_1}$ and $\varphi_{S_2, \lambda_2}$ are quasi-equivalent.
The quasi-equivalence of $\varphi_{S_1, \lambda_1}$ and $\varphi_{S_2, \lambda_2}$ induces the quasi-equivalence of $\varphi_{S_1, \lambda_1 - \lambda_2}$ and $\varphi_{S_2}$.
Put $\lambda := \lambda_1 - \lambda_2$.
Then there exists a $*$-isomorphism $\tau$ from $\pi_{S_1, \lambda}(\calW(V, \sigma))^\dprime$ onto $\pi_{S_2}(\calW(V, \sigma))^\dprime$ such that
\begin{equation}
\tau(\pi_{S_1, \lambda}(A)) = \pi_{S_2}(A), \quad A \in \calW(V, \sigma). \label{eq:qe iso q-f state and coh state}
\end{equation}
For any $f \in V$, 
\begin{eqnarray}
& & \exp(i \lambda(f) - S_1(f,f)/2) = \innpro{\xi_{S_1}, \tau^{-1}(\pi_{S_2}(W(f)))  \xi_{S_1} } \nonumber\\
&=& \innpro{\xi_{S_1}, \tau^{-1}(\pi_{S_2}(W(f)))  \xi_{S_1} }  \label{eq:continuity of each other}
\end{eqnarray}
is $\norm{\cdot}_{S_2}$-continuous in $f \in V$.
Thus, $\lambda$ and $S_1$ are $\norm{\cdot}_{S_2}$-continuous.
By symmetry, $\lambda$ and $S_2$ are $\norm{\cdot}_{S_1}$-continuous as well.
By Lemma \ref{lmm:kernel property}, $N_S := N_{S_1} = N_{S_2}$.
If $\lambda \neq 0$ on $N_S$, then there exists $f \in N_S \backslash \{ 0\}$ such that $\lambda(f) \neq 0$.
If $\lambda(f) = 2n\pi$ for some $n \in \bbZ$, then we replace $f$ by $f/\pi$.
For such $f$, we have 
\begin{equation}
e^{i \lambda(f)} = \tau(\pi_{S_1, \lambda}(W(f))) = \pi_{S_2}(W(f)) = \mathbbm{1} \label{eq:contradiction q-e state and coh state}
\end{equation}
by Lemma \ref{lmm:kernel property}.
It contradicts to the quasi-equivalence of $\varphi_{S_1, \lambda}$ and $\varphi_{S_2}$.
Thus, $\lambda = 0$ on $N_S$.
Let $\tau^\prime$ be the map from $\pi_{S_1, \lambda}(\calW(V, \sigma))$ to $\pi_{S_1}(\calW(V, \sigma))$ defined by
\begin{equation}
\tau^\prime(\pi_{S_1, \lambda}(A)) = \pi_{S_1}(A), \quad A \in \calW(V, \sigma). \label{eq:qe iso 2 q-f state and coh state}
\end{equation}
Since $\lambda$ is continuous with respect to the norm $\norm{\cdot}_{S_1}$ and $\lambda = 0$ on $N_S$, then we can extend $\tau^\prime$ to a map from $\pi_{S_1, \lambda}(\calW(V, \sigma))^\dprime$ onto $\pi_{S_1}(\calW(V, \sigma))^\dprime$.
Then $\tau^\prime$ induce the quasi-equivalence of $\varphi_{S_1, \lambda}$ and $\varphi_{S_1}$.
Thus, $\varphi_{S_1}$ and $\varphi_{S_2}$ are quasi-equivalent and by Theorem \ref{theorem:ArakiYamagami}, we have the statement.
\QED

\begin{rmk}
In {\rm \cite{Yamagami12}}, S. Yamagami obtained quasi-equivalence conditions of (generalized) coherent states in terms of the transition amplitude.
For applications to concrete models Hilbert-Schmidt conditions in Theorem {\rm \ref{theorem:quasi-equivalence of coherent states}} are easier to handle.
Let $\varphi_{S_1, \lambda_1}$ and $\varphi_{S_2, \lambda_2}$ be generalized coherent states on the Weyl CCR algebra $\calW(V, \sigma)$.
Assume that $\varphi_{S_1}$ and $\varphi_{S_2}$ are quasi-equivalent.
If $\lambda_1 - \lambda_2$ is not continuous in $\norm{\cdot}_{S_1}$ or $\norm{\cdot}_{S_2}$ or $\lambda_1 \neq \lambda_2$, then the transition amplitude $(\varphi_{S_1, \lambda_1}^{1/2}, \varphi_{S_2, \lambda_2}^{1/2}) = 0$, where $\varphi_{S_1}^{1/2}$ and $\varphi_{S_2, \lambda_2}^{1/2}$ is GNS-vector in the universal representation space $L^2(\calW(V, \sigma)^{**})$.
{\rm (See \cite[Theorem 5.3.]{Yamagami12}.)}
\end{rmk}

Factor decompositions of quasi-free states are given in \cite{Honegger90}, \cite{RoccaSirugueTestard} and \cite{Yamagami10}, e.t.c..
For the convenience of the reader, we give an explicit form of factor decomposition of a non-factor generalized coherent state.
We recall the definition of the disjointness of states.
(See also \cite[Definition 4.1.20.]{BratteliRobinsonI} and \cite[Lemma 4.2.8.]{BratteliRobinsonI}.)
\begin{df} \label{df:disjointness}
Let $\varphi_1$ and $\varphi_2$ be positive linear functionals on a ${\rm C}^*$-algebra $\calA$.
The positive linear functionals $\varphi_1$ and $\varphi_2$ are said to be disjoint, if for $\omega = \varphi_1 + \varphi_2$, there is a projection $P \in \pi_\omega(\calA)^\dprime \cap \pi_\omega(\calA)^\prime$ such that 
\begin{eqnarray}
\varphi_1(A) &=& (\xi_\omega, P\pi_\omega(A) \xi_\omega), \nonumber\\
\varphi_2(A) &=& (\xi_\omega, (\mathbbm{1} - P) \pi_\omega(A) \xi_\omega), \quad A \in \calA,
\end{eqnarray}
where $\pi_\omega$ is the GNS-representation and $\xi_\omega$ is the GNS-cyclic vector associated with $\omega$.
\end{df}

Note that factor representations are either quasi-equivalent or disjoint.
(See e.g. \cite[Proposition 2.4.22.]{BratteliRobinsonI}, \cite[Theorem 2.4.26. (1)]{BratteliRobinsonI}, and \cite[Proposition 2.4.27.]{BratteliRobinsonI}.)

\begin{theorem} \label{theorem:factor decomp of generalized coherent state}
Let $\varphi_{S, \lambda}$ be a generalized coherent state on $\calW(V, \sigma)$.
If $\varphi_{S, \lambda}$ is non-factor, then there exists a probability measure $\mu$ on $\bbR^{2I}$ and $\varphi_{S, \lambda}$ has factor decomposition of the form
\begin{equation}
\varphi_{S, \lambda} = \int_{\bbR^I} \varphi_{S E_0^\perp, x\cdot \rho + \lambda} d\mu(x),
\end{equation}
where $\varphi_{S E_0^\perp, x \cdot \rho + \lambda}(W(f)) = \exp(- S(E_0^\perp f, E_0^\perp f)/4 + i x \cdot \rho(f) + i \lambda(f))$ and $\rho(f) = ({\rm Re}(e_k,f)_S, {\rm Im}(e_k, f)_S)_{k \in I} \in \bbR^{2I}$.
Moreover, $\varphi_{S E_0^\perp, x \cdot \rho + \lambda}$ and $\varphi_{S E_0^\perp, y \cdot \rho + \lambda}$ are disjoint unless $x \neq y$, $x, y \in \bbR^{2I}$.
\end{theorem}

\noindent {\bf Proof.}
If a generalized coherent state $\varphi_{S, \lambda}$ on $\calW(V, \sigma)$ is non-factor, then on $V^\bbC_S$, $S$ has the spectral decomposition 
\begin{equation}
S f = S E_0^\perp f + \frac{1}{2} \sum_{k \in I} (e_k, f)_S e_k, \quad f \in V^\bbC_S, \label{eq:decomposition of S}
\end{equation}
where $E_0$ is the spectral projection of $S$ with an eigenvalue $1/2$, $I$ is an index set such that $\abs{I} = {\rm dim}\ker(S - 1/2)$, and $\{ e_k \}_{k \in I}$ is an orthonormal basis for $\ker(S - 1/2)$.
Thus, for any $W(f)$, $f \in V$, we have
\begin{equation}
\varphi_{S, \lambda}(W(f)) = \exp(- \frac{S(E_0^\perp f, E_0^\perp f)}{4} + i \lambda(f) ) \exp(- \frac{\sum_{k \in I} \abs{(e_k, f)_{S}}^2}{8}). \label{eq:decomposition of non-factor generalized coherent state}
\end{equation}
By a theorem of Bochner--Minlos (See e.g. \cite[Theorem 2.2.]{Simon}), there exists a probability measure $\mu$ on $\bbR^{2I}$ such that 
\begin{equation}
\exp(- \frac{\sum_{k} \abs{(e_k, f)_S}^2}{8}) = \int_{\bbR^{2I}} \exp(i x \cdot \rho(f)) d\mu(x), \label{eq:factor decomposition of non-gauge invariant generalized coherent state} 
\end{equation}
where $\rho(f) = ( {\rm Re} (e_k, f)_S, {\rm Im} (e_k, f)_{S} )_{k \in I} \in \bbR^{2I}$.
For $\varphi_{SE_0^\perp, x \cdot \rho + \lambda}$, we have $N_{S E_o^\perp} = E_0 V^\bbC \neq \0$.
Since $E_0 V^\bbC \neq \0$, there exists a $f \in V$ such that ${\rm Re}(e_k, f)_S \neq 0$ or ${\rm Im}(e_k, f)_S \neq 0$.
We put $f_n := E_0 f + 1/n E_0^\perp f$.
Then $\norm{f_n}_{SE_0^\perp} \to 0$ and ${\rm Re}(e_k, f_n)_S \not \to 0$ or ${\rm Im}(e_k, f_n)_S \not \to 0$ as $n \to \infty$.
Thus, the generalized coherent states $\varphi_{S E_0^\perp, x \cdot \rho + \lambda}$ and $\varphi_{S E_0^\perp, y \cdot \rho + \lambda}$, $x, y \in \bbR^{2I}$ are not quasi-equivalent unless $x = y$ by Theorem \ref{theorem:quasi-equivalence of coherent states}.
Since $\norm{\cdot}_{S}$ and $\norm{\cdot}_{SE_0^\perp}$ induce the same topology on $V^\bbC$ and $SE_0^\perp$ on $V^\bbC_{SE_0^\perp}$ does not have an eigenvalue $1/2$, $\varphi_{S E_0^\perp, x \cdot \rho + \lambda}$ is factor and $\varphi_{S E_0^\perp, x \cdot \rho + \lambda}$ and $\varphi_{S E_0^\perp, y \cdot \rho + \lambda}$ are disjoint unless $x \neq y$, $x, y \in \bbR^{2I}$.
\QED

%%%%%%%%%%%%%%%%%%%%%%%%%%%%%%%%%%%%%%%%%%%%%%%%%%
%%%%%%%%%%%%%%%%%%%%%%%%%%%%%%%%%%%%%%%%%%%%%%%%%%
%%%%%%%%%%%%%%%%%%%%%%%%%%%%%%%%%%%%%%%%%%%%%%%%%%
%%%%%%%%%%%%%%%%%%%%%%%%%%%%%%%%%%%%%%%%%%%%%%%%%%
%%%%%%%%%%%%%%%%%%%%%%%%%%%%%%%%%%%%%%%%%%%%%%%%%%
%%%%%%%%%%%%%%%%%%%%%%%%%%%%%%%%%%%%%%%%%%%%%%%%%%
%%%%%%%%%%%%%%%%%%%%%%%%%%%%%%%%%%%%%%%%%%%%%%%%%%
%%%%%%%%%%%%%%%%%%%%%%%%%%%%%%%%%%%%%%%%%%%%%%%%%%
%%%%%%%%%%%%%%%%%%%%%%%%%%%%%%%%%%%%%%%%%%%%%%%%%%
%%%%%%%%%%%%%%%%%%%%%%%%%%%%%%%%%%%%%%%%%%%%%%%%%%
%%%%%%%%%%%%%%%%%%%%%%%%%%%%%%%%%%%%%%%%%%%%%%%%%%
%%%%%%%%%%%%%%%%%%%%%%%%%%%%%%%%%%%%%%%%%%%%%%%%%%
%%%%%%%%%%%%%%%%%%%%%%%%%%%%%%%%%%%%%%%%%%%%%%%%%%
%%%%%%%%%%%%%%%%%%%%%%%%%%%%%%%%%%%%%%%%%%%%%%%%%%
%%%%%%%%%%%%%%%%%%%%%%%%%%%%%%%%%%%%%%%%%%%%%%%%%%
%%%%%%%%%%%%%%%%%%%%%%%%%%%%%%%%%%%%%%%%%%%%%%%%%%
%%%%%%%%%%%%%%%%%%%%%%%%%%%%%%%%%%%%%%%%%%%%%%%%%%
%%%%%%%%%%%%%%%%%%%%%%%%%%%%%%%%%%%%%%%%%%%%%%%%%%
%%%%%%%%%%%%%%%%%%%%%%%%%%%%%%%%%%%%%%%%%%%%%%%%%%
%%%%%%%%%%%%%%%%%%%%%%%%%%%%%%%%%%%%%%%%%%%%%%%%%%
%%%%%%%%%%%%%%%%%%%%%%%%%%%%%%%%%%%%%%%%%%%%%%%%%%
%%%%%%%%%%%%%%%%%%%%%%%%%%%%%%%%%%%%%%%%%%%%%%%%%%
%%%%%%%%%%%%%%%%%%%%%%%%%%%%%%%%%%%%%%%%%%%%%%%%%%
%%%%%%%%%%%%%%%%%%%%%%%%%%%%%%%%%%%%%%%%%%%%%%%%%%
\setcounter{theorem}{0}
\setcounter{equation}{0}
\section{BEC and Non-factor states} \label{sec:graphs}
In this section, we consider quasi-free states on $\calW(\frah, \sigma)$, where $\frah$ is a pre-Hilbert space over $\bbC$ with an inner product $\innpro{\cdot, \cdot}_{\frah}$ and $\sigma(f,g) = {\rm Im}\innpro{f,g}_{\frah}$, $f,g \in \frah$.
We give the decomposition of quasi-free states on $\calW(\frah, \sigma)$ into generalized coherent states which are mutually disjoint.

\subsection{General properties}
In this subsection, we use the following notations.
Let $\frah$ be a subspace of a Hilbert space over $\bbC$.
We assume that $\frah$ is equipped with positive definite inner products $\innpro{\cdot, \cdot}_{\frah}$ and $\innpro{\cdot, \cdot}_{0}$.
Let $q$ be a linear functional on $\frah$.
We consider the quasi-free state $\varphi_{q,D}$, $D \geq0$, on $\calW(\frah, \sigma)$ defined by
\begin{equation}
\varphi_{q,D} (a^\dagger(f) a(g)) = \innpro{ g, f }_{0} + D\overline{q(g)}q(f), \label{eq:two-point functional}
\end{equation}
where $a(f)$ and $a^\dagger(f)$, $f \in \frah$, are the annihilation operators and the creation operators on the GNS representation space $\fraH_{\varphi_{q,D}}$, respectively.
Note that the annihilation operators $a(f)$, $f \in \frah$, and the creation operators $a^\dagger(f)$, $f \in \frah$ satisfy the following equation:
\begin{equation}
\sbk{ a(f), a^\dagger(g) } = \innpro{f,g}_{\frah}, \quad \sbk{a(f), a(g)} = 0 = \sbk{a^\dagger(f), a^\dagger(g)}, \quad f,g \in \frah.
\end{equation}

Our aim is to show that $\varphi_{q, D}$ is non-factor if $q$ is not continuous with respect to the norm $\norm{ \cdot }_{q,D}$ defined in (\ref{eq:innpro of KD}) and $D > 0$, and to get factor decomposition of $\varphi_{q,D}$, in this subsection.
Let $\{ e_n \}_{n \in \bbN}$ be an orthonormal basis on a Hilbert space which is contained in $\frah$.
Fix $\{e_n\}_{n \in \bbN}$.
We set 
\begin{equation}
\overline{f} = \sum_{n \in \bbN} \overline{f_n} e_n, \label{eq:complex conjugate}
\end{equation}
for $f = \sum_{n \in \bbN} f_n e_n \in \frah$, where $f_n \in \bbC$, $n \in \bbN$ and $\overline{f_n}$ is the complex conjugate of $f_n$.
For a linear functional $q$ and $D \geq 0$, we put $\tilde{K}_{q,D} = \frah \oplus \frah$.
For $f_1, f_2, g_1, g_2 \in \frah$, we sets
\begin{eqnarray}
& & \gamma_D(f_1 \oplus f_2, g_1 \oplus g_2) = \frac{1}{2} \rbk{\innpro{f_1, g_1}_{\frah} - \innpro{f_2, g_2}_{\frah}}, \label{eq:gammah}\\
& & \Gamma(f_1 \oplus f_2) = \overline{f_2} \oplus \overline{f_1}, \label{eq:Gammah}\\
& & B(f_1 \oplus f_2) = \frac{1}{\sqrt{2}} \rbk{a^\dagger(f_1) + a(\overline{f_2})}, \label{eq:Bh}\\
& & S_{q,D}(f_1 \oplus f_2, g_1 \oplus g_2) = \varphi_{q,D}(B(f_1 \oplus f_2)^*B(g_1 \oplus g_2)) \nonumber\\
&=& \frac{1}{2} \varphi_{q,D}((a^\dagger(f_1) + a(\overline{f_2}))^*(a^\dagger(g_1) + a(\overline{g_2}))) \nonumber\\
&=& \frac{1}{2} \innpro{f_1,g_1}_{\frah} + \frac{1}{2} \innpro{f_1, g_1}_{0} + \frac{1}{2} \innpro{f_2, g_2}_{0} + \frac{D}{2} \overline{q(f_1)} q( g_1) + \frac{D}{2} \overline{q(f_2)} q( g_2).\label{eq:Sh}
\end{eqnarray}
We define the inner product on $\tilde{K}_{q,D}$ by
\begin{eqnarray}
\innpro{f_1 \oplus f_2, g_1 \oplus g_2}_{q,D} &=& \frac{1}{2} \innpro{f_1,g_1}_\frah + \frac{1}{2} \innpro{f_2, g_2}_\frah + \innpro{f_1, g_1}_{0} + \innpro{f_2, g_2}_{0} \nonumber\\
& & + D \overline{q(f_1)} q(g_1) + D \overline{q(f_2)} q( g_2). \label{eq:innpro of KD}
\end{eqnarray}
Let $N_{K_{q,D}} = \set{f \in \tilde{K}_{q,D} | \norm{f}_{q,D} = 0}$.
Then we denote the completion of $\tilde{K}_{q,D} / N_{K_{q,D}}$ with respect to the norm $\norm{ \cdot }_{q,D}$ by $K_{q,D}$.
In this case, $\norm{f_1 \oplus f_2}_{q,D} = 0$ leads $f_1 = 0$ and $f_2 = 0$.
Thus, $N_{K_{q,D}} = \0$.

We put
\begin{equation}
\innpro{f, g}_{\fraK} = \frac{1}{2} \innpro{f,g}_{\frah} + \innpro{f, g}_{0}, \quad f,g \in \frah, \label{eq:inn pro fraK}
\end{equation}
and $\norm{\cdot}_{\fraK} = \innpro{\cdot, \cdot}_{\fraK}^{1/2}$.
We define the Hilbert space $\fraK$ by the completion of $\frah$ with respect to the norm $\norm{\cdot}_{\fraK}$.

\begin{lmm} \label{lmm:isomorphich}
The space $K_{q,D}$ has the following form:
\begin{itemize}
\item[$1.$] If $D>0$ and $q$ is not continuous with respect to the norm $\norm{\cdot}_{\fraK}$, then we have
\begin{equation}
K_{q,D} = \bbC \oplus \fraK \oplus \bbC \oplus \fraK, \label{eq:isomorphic to K_D}
\end{equation}
\item[$2.$] If $D = 0$ or $q$ is continuous with respect to the norm $\norm{\cdot}_{\fraK}$, then we have
\begin{equation}
K_{q,D} = \fraK \oplus \fraK. \label{eq:isomorphic to K_D,D=0}
\end{equation}
\end{itemize}
\end{lmm}

\noindent {\bf Proof.}
We consider the case of $D > 0$ and $q$ is not continuous with respect to the norm $\norm{\cdot}_{\fraK}$.
It suffices to show that $\bbC \oplus \fraK = \overline{\frah}$, where $\overline{\frah}$ is the completion of $\frah$ with respect to the norm $\norm{ \cdot }^\prime$ defined by
\begin{equation}
(\norm{f}^\prime)^2 = \norm{f}_{\fraK}^2 + D \abs{q(f)}^2, \quad f \in \frah \label{eq:half norm h}
\end{equation}
We define $\pi : \frah \to \bbC \oplus \fraK$ by
\begin{equation}
\pi(f) = q( f ) \oplus f. \label{eq:inclusion map h}
\end{equation}
Since $q$ is not continuous, for any $f \in \frah$, there exists a sequence $f_n$ in $\frah$ such that $\lim_{n \to \infty} \norm{f_n - f}_{\fraK} = 0$ and $\lim_{n \to \infty}q(f_n) = 0$.
For such $f_n$ and $f$, we have
\begin{equation}
\pi(f_n - f) \to q(f) \oplus 0, \quad \pi(f_n) \to 0 \oplus f. \label{eq:inclusion limit h}
\end{equation}

The case of $D = 0$ is clear. 
We assume that $q$ is continuous with respect to the norm $\norm{\cdot}_{\fraK}$.
By continuity of $q$, the norm $\norm{\cdot}^\prime$, defined in (\ref{eq:half norm h}), and $\norm{\cdot}_{\fraK}$ induce the same topology.
\QED

\begin{theorem} \label{theorem:non-factor}
For a linear space $\frah$ with positive definite inner products $\innpro{\cdot, \cdot}_{\frah}$ and $\innpro{\cdot, \cdot}_{0}$, if $D > 0$ and $q$ is not continuous with respect to the norm $\norm{\cdot}_\fraK$, then the two-point function $\varphi_{q,D}$ defined in {\rm (\ref{eq:two-point functional})} is a non-factor state on $\calW(\frah, \sigma)$.
\end{theorem}

\noindent {\bf Proof.}
First, we consider the case of $D > 0$.
By Lemma \ref{lmm:araki1} and Lemma \ref{lmm:araki2}, it suffices to show that $1/2 \in \sigma_P(S_{q,D})$.
By Lemma \ref{lmm:isomorphich}, an element of $K_ {q, D}$ has the form $(a_1, f_1, a_2, f_2)$, $a_1, a_2 \in \bbC$, $f_1, f_2 \in \fraK$.
For any $(a_1 \oplus f_1 \oplus a_2 \oplus f_2), (b \oplus 0 \oplus 0 \oplus 0) \in K_{q,D}, b \in \bbC$, the operator $S_{q,D}$ satisfies 
\begin{eqnarray}
& & \innpro{ (a_1 \oplus f_1 \oplus a_2 \oplus f_2), S_{q,D} (b_1 \oplus 0 \oplus 0 \oplus 0) }_{q, D} = \frac{D}{2}\overline{a_1} b \nonumber \\
&=& \frac{1}{2} \innpro{ (a_1 \oplus f_1 \oplus a_2 \oplus f_2), (b \oplus 0 \oplus 0 \oplus 0) }_{q, D}. \label{eq:realization of S}
\end{eqnarray}
Thus, we have $S_{q,D}(b \oplus 0 \oplus 0 \oplus 0) = 1/2(b \oplus 0 \oplus 0 \oplus 0)$ for any $b \in \bbC$ and $1/2 \in \sigma_P(S_{q,D})$.
\QED

\begin{pro} \label{pro:factor}
For a linear space $\frah$ with positive definite inner products $\innpro{\cdot, \cdot}_{\frah}$ and $\innpro{\cdot, \cdot}_{0}$, if $D = 0$ or $q$ is continuous with respect to the norm $\norm{\cdot}_\fraK$, the two-point function $\varphi_{q,D}$ defined in {\rm (\ref{eq:two-point functional})} is a factor state on $\calW(\frah, \sigma)$.
\end{pro}

\noindent {\bf Proof.}
If $q$ is continuous with respect to the norm $\norm{\cdot}_\fraK$, then $\varphi_{q,D}$ is quasi-equivalent to $\varphi_{0,0}$ by Theorem \ref{theorem:quasi-equivalence of coherent states}.
Thus, it suffice to show the case of $D = 0$.
There exists the positive contraction operator $A$ on $\fraK$ such that $\innpro{\xi, A \eta}_{\fraK} = \innpro{\xi, \eta}_{\frah}/2$ and $\innpro{\xi, (\mathbbm{1} - A) \eta}_{\fraK} = \innpro{\xi, \eta}_0$, $\xi, \eta \in \fraK$.
Then $S_{0,0}$ has the following form:
\begin{equation}
S_{0,0}(\eta_1 \oplus \eta_2) = (A + (\mathbbm{1} - A)/2) \eta_1 \oplus \frac{\mathbbm{1} - A}{2} \eta_2 = \frac{\mathbbm{1} + A}{2} \eta_1 \oplus \frac{\mathbbm{1} - A}{2} \eta_2,
\end{equation}
for $\eta_1, \eta_2 \in \fraK$.
If $1/2 \in \sigma_P(S_{0,0})$, then $(\mathbbm{1} + A) \eta_1 = \eta_1$ and $(\mathbbm{1} - A) \eta_2 = \eta_2$.
Thus, $\eta_1, \eta_2 \in \ker A$.
Since the positive definiteness of $\innpro{\cdot, \cdot}_\frah$ and $\innpro{\cdot, \cdot}_0$ on $\frah$, $\frah \cap \ker A = \0$.
Thus, $\ker A = \0$ and $\varphi_{0,0}$ is factor.
\QED

Next, we consider factor decomposition of $\varphi_{q,D}$, if $q$ is not continuous in $\norm{\cdot}_\fraK$.
Let $(\fraH_0, \pi_0, \xi_0)$ be the GNS-representation space with respect to $\varphi_0 := \varphi_{q, 0} = \varphi_{0, D}$.
Since $\varphi_0$ is regular state on $\calW(\frah, \sigma)$, there exist self-adjoint operators $\Psi_0(f)$, $f \in \frah$, such that
\begin{equation}
\pi_0(W(f)) = \exp(i \Psi_0(f)). \label{eq:Weyl generator}
\end{equation}
Now we define the field operators $\Psi_{s_1, s_2}(f)$, $s_1, s_2 \in \bbR$, $f \in \frah$, on $\fraH_0$ by
\begin{equation}
\Psi_{s_1, s_2}(f) = \Psi_0(f) + s_1D^{1/2} {\rm Re} q(f) \mathbbm{1} + s_2 D^{1/2} {\rm Im} q(f) \mathbbm{1}, \quad f \in \frah. \label{eq:def of psi_{s_1, s_2}} 
\end{equation}
Let $\pi_{s_1, s_2}$ be the representation of $\calW(\frah, \sigma)$ on $\fraH_0$ defined by
\begin{equation}
\pi_{s_1, s_2}(W(f)) = \exp(i \Psi_{s_1, s_2}(f)), \quad f \in \frah. \label{eq:def of repn s_1, s_2}
\end{equation} 
Using the $\pi_{s_1, s_2}$, we define the state $\varphi_{s_1, s_2}$ on $\calW(\frah, \sigma)$ by
\begin{equation}
\varphi_{s_1, s_2}(A) = \innpro{ \xi_0, \pi_{s_1, s_2} (A) \xi_0}, \quad A \in \calW(\frah, \sigma). \label{eq:def of state s_1, s_2}
\end{equation}
Then we have the following theorem.

\begin{theorem} \label{theorem:disjointness}
If $q$ is not continuous in $\norm{\cdot}_\fraK$, then for each $s_1, s_2, t_1, t_2 \in \bbR$, $\varphi_{s_1, s_2}$ and $\varphi_{t_1, t_2}$ are factor and disjoint unless $t_1 = s_1$ and $t_2 = s_2$.
\end{theorem}

\noindent {\bf Proof.}
By Lemma \ref{lmm:equality of von Neumann algebras} and Proposition \ref{pro:factor}, $\varphi_{s_1, s_2}$ and $\varphi_{t_1, t_2}$ are factor.
Since $q$ is not continuous with respect to the norm, $\varphi_{s_1, s_2}$ and $\varphi_{t_1, t_2}$ are disjoint unless $t_1 = s_1$ and $t_2 = s_2$ by Theorem \ref{theorem:quasi-equivalence of coherent states}.
\QED

Finally, we obtain factor decomposition of $\varphi_{q, D}$.

\begin{theorem} \label{theorem:factor decomposition}
If $q$ is not continuous in $\norm{\cdot}_\fraK$, then for any $D > 0$, factor decomposition of $\varphi_{q, D}$ defined in {\rm (\ref{eq:two-point functional})} is given by
\begin{equation}
\varphi_{q, D} = \frac{1}{2 \pi} \int_{\bbR^2} \varphi_{s_1, s_2} e^{- \frac{s_1^2 + s_2^2}{2}} ds_1 ds_2. \label{eq:factor decomp}
\end{equation}
\end{theorem}

\noindent {\bf Proof.}
By Theorem \ref{theorem:factor decomp of generalized coherent state}, we are done.
\QED

\subsection{On graphs}
In this subsection, let $X = (VX, EX)$ be an undirected graph, where $VX$ is the set of all vertices in $X$ and $EX$ is the set of all edges in $X$.
Two vertices $x, y \in VX$ are said to be adjacent if there exists an edge $(x, y) \in EX$ joining $x$ and $y$, and we write $x \sim y$.
We denote the set of all the edges connecting $x$ with $y$ by $E_{x,y}$.
Since the graph is undirected, $E_{x,y} = E_{y,x}$.
Let $\ell^2(VX)$ be the set of all square summable sequence labeled by the vertices in $VX$.
Let $A_X$ be the adjacency operator of $X$ defined by
\begin{equation}
\innpro{\delta_x, A_X \delta_y} = \abs{E_{x, y}}, \quad x,y \in VX. \label{eq:def of adjacency}
\end{equation}
In addition, for any $x \in VX$, we set the degree of $x$ by ${\rm deg}(x)$ and 
\begin{equation}
{\rm deg} := \sup_{x \in VX} {\rm deg}(x). \label{eq:def of degree of graph}
\end{equation}
We assume that $X$ is connected, countable and ${\rm deg} < \infty$.
Then, the adjacency operator $A_X$ acting on $\ell^2(VX)$ is bounded.
If for any $\delta_x$, $x \in VX$, $A_X$ satisfies the condition
\begin{equation}
\lim_{\lambda \searrow \norm{A_X}} \innpro{ \delta_x, (\lambda \mathbbm{1} - A_X)^{-1} \delta_x } < \infty, \label{eq:transience}
\end{equation}
then $A_X$ is said to be transient.
Let $H$ be the Hamiltonian on $\ell^2(VX)$ defined by $H := \norm{A_X} \mathbbm{1} - A_X$.

A bounded operator $B$ on $\ell^2(VX)$ is called positivity preserving if $B_{x,y} := \innpro{\delta_x, B \delta_y} \geq 0$ for any $x, y \in VX$.
A sequence $\set{ v(x) | x \in VX }$ is called a Perron--Frobenius weight for $B$ if it has positive entries and 
 \begin{equation}
\sum_{y \in VX} B_{x,y} v(y) = \norm{B} v(x) \label{eq:Perron--Frobenius weight}
\end{equation}
for any $x \in VX$.

In \cite{Fidaleo15}, F. Fidaleo considered BEC on graphs and showed the following two results.

\begin{pro} {\rm \cite[Proposition 4.1.]{Fidaleo15}} \label{pro:fidaleo1}
Let $A_X$ be the adjacency operator of $X$ on $\ell^2(VX)$ and $H$ be the Hamiltonian defined by $H = \norm{A_X} \mathbbm{1} - A_X$.
Let $\frah$ be a subspace of $\ell^2(VX)$ satisfying the following three conditions:
For each $\beta > 0$,
\begin{enumerate}
\item[$1.$] $e^{itH} \frah = \frah$, $t \in \bbR$;
\item[$2.$] For each entire function $f$, $f(H) \frah \subset \calD( (e^{\beta H} - 1)^{-1/2} )$;
\item[$3.$] $\sum_{x \in VX} \abs{ (f(H) u)(x) } v(x) < \infty$, and $\innpro{ f(H) u, v } = \overline{f(0)} \innpro{u, v}$, where $v$ is a Perron--Frobenius weight for $A_X$.
\end{enumerate}
Then for $D \geq 0$, the two--point function
\begin{equation}
\varphi_D(a^*(f_1)a(f_2)) = \innpro{ (e^{\beta H} - \mathbbm{1})^{-1} f_2, f_1 }_{\ell^2} + D\innpro{f_2, v} \innpro{v, f_1} \label{eq:BECoccur}
\end{equation}
satisfies the KMS condition at inverse temperature $\beta > 0$ on the Weyl CCR algebra $\calW(\frah, \sigma)$ with respect to the dynamics generated by the Bogoliubov transformations
\begin{equation}
f \in \frah \mapsto e^{itH} f, \quad t \in \bbR. \label{eq:Bogoliubov transformation}
\end{equation}
\end{pro}

By the above proposition and \cite[Proposition 1.1.]{Matsui}, we are said to be BEC occur if the case of $D > 0$ and BEC does not occur if the case of $D = 0$.

\begin{theorem} {\rm \cite[Theorem 4.5.]{Fidaleo15}} \label{th:fidaleo2}
Suppose that $A_X$ is transient.
Let $\frah_1$ be the subspace of $\ell^2(VX)$ defined by
\begin{equation}
\frah_1 = \set{ e^{itH} \delta_x | t \in \bbR, x \in VX }. \label{eq:def of subspace1}
\end{equation}
Then $\frah_1$ satisfies the conditions $1$, $2$, and $3$ in Proposition {\rm \ref{pro:fidaleo1}}.
Thus, for $\frah_1$ and any $D \geq 0$, the two-point function given in {\rm (\ref{eq:BECoccur})} defines KMS state on the Weyl CCR algebra $\calW(\frah_1, \sigma)$. 
\end{theorem}

We give another example of $\frah$.
Let $\calP(\bbC)$ be the set of all polynomial functions on $\bbC$.
Let $\frah_2$ be the subspace defined by 
\begin{equation}
\frah_2 = \set{ \int_\bbR p(t) e^{-(t - a)^2/b} e^{itH} \delta_x dt | p \in \calP(\bbC), a \in \bbR, b > 0, x \in VX }. \label{eq:def of subspace2}
\end{equation}

\begin{lmm} \label{lmm:frah satisfies three conditions}
The space $\frah_2$ satisfies the following conditions;
\begin{enumerate}
\item[$1^\prime$.] $e^{itH} \frah_2 = \frah_2$, $t \in \bbR$;
\item[$2^\prime$.] $e^{\beta H} \frah_2 \subset \calD((e^{\beta H} - 1)^{-1/2})$;
\item[$3^\prime$.] $\sum_{x \in VG} \abs{ (e^{\beta H} u)(x) } < \infty$, and $\innpro{e^{\beta H} u, v} = \innpro{ u,v }$, $u \in \frah_2$.
\end{enumerate}
\end{lmm}

\noindent {\bf Proof.}
The condition $1^\prime$, $e^{itH} \frah_2 \subset \frah_2$ is clear.
Now we prove the condition $2^\prime$, $e^{\beta H} \frah_2 \subset \calD((e^{\beta H} - \mathbbm{1})^{-1/2})$.
Note that $(e^{\beta x} - 1)^{-1} - (\beta x)^{-1}$ is continuous on $[0, \infty)$.
Thus, it enough to show that $e^{\beta H} \frah_2 \subset \calD(H^{-1/2})$.
Since $A_X$ is transient and $p(t) e^{-(t - a)^2/b}$ is a rapidly decreasing function on $\bbR$, for a generator of $\frah_2$, $\int_\bbR p(t) e^{- \frac{(t-a)^2}{b}} e^{itH} \delta_x dt$, we have
\begin{eqnarray}
& & \innpro{ (\lambda \mathbbm{1} - A_X)^{-1} e^{\beta H} \int_\bbR p(t) e^{- \frac{(t-a)^2}{b}} e^{itH} \delta_x dt, e^{\beta H} \int_\bbR p(t) e^{- \frac{(t-a)^2}{b}} e^{itH} \delta_x dt} \nonumber\\
&=& \abs{ \int_\bbR \int_\bbR \overline{p(t)} p(s) e^{- \frac{(t-a)^2}{b}} e^{- \frac{(s-a)^2}{b}} \innpro{ (\lambda\mathbbm{1} - A_X)^{-1} e^{\beta H} e^{itH} \delta_x, e^{\beta H} e^{isH} \delta_x } dtds} \nonumber\\
&=& \abs{ \int_\bbR \int_\bbR \overline{p(t)} p(s) e^{- \frac{(t-a)^2}{b}} e^{- \frac{(s-a)^2}{b}} \int_{\sigma(A_X)} \frac{e^{i(s - t) a}  e^{ 2\beta (\norm{A_X} \mathbbm{1} - a)} } {\lambda - a} d\innpro{\delta_x, E(a) \delta_x} dt ds} \nonumber \\
&\leq& C_1  e^{4 \beta \norm{A_X} } \innpro{(\lambda \mathbbm{1} - A_X)^{-1} \delta_x, \delta_x} \nearrow C_1 e^{2\beta \norm{A_X}}\innpro{(\norm{A_X} \mathbbm{1} - A_X)^{-1} \delta_x, \delta_x} < \infty, \nonumber\\
\label{eq:generator in the domain of inverse of H}
\end{eqnarray}
where $C_1$ is a positive constant satisfying
\begin{equation}
  \int_\bbR \int_\bbR \abs{ \overline{p(t)} p(s) e^{- \frac{(t-a)^2}{b}} e^{- \frac{(s-a)^2}{b}} } dt ds < C_1.
\end{equation}
Next, we show that $\sup_{n \in \bbN} \sum_{x \in V\Lambda} \abs{ (e^{\beta H} u)(x) } v (x) < \infty$, $u \in \frah_2$, where $\Lambda_n$ is a finite subgraph of $X$ such that $\Lambda_n \nearrow X$.
Let $C_R$ be a circle centered at the origin with radius $R > \norm{A_X}$.
We have
\begin{eqnarray}
& & \abs{ \innpro{ e^{\beta H} \int_\bbR p(t) e^{- \frac{(t-a)^2}{b}} e^{itH} \delta_x dt, \delta_y} } \leq \int_\bbR \abs{ p(t) } e^{- \frac{(t-a)^2}{b}} \abs{ \innpro{ e^{\beta H} e^{itH} \delta_x, \delta_y } } dt \nonumber\\
&=& \int_\bbR \abs{ p(t) } e^{- \frac{(t-a)^2}{b}} \abs{ \frac{1}{2\pi i} \oint_{C_R} e^{\beta z} e^{it z} \innpro{ (z \mathbbm{1} - A_X)^{-1} \delta_x, \delta_y }dz } dt \nonumber\\
&\leq& R e^{\beta R} \int_\bbR \abs{ p(t) } e^{- \frac{(t-a)^2}{b}}  e^{t R} dt \innpro{ (R \mathbbm{1} - A_X)^{-1} \delta_x, \delta_y } \leq C_2 \innpro{ (R \mathbbm{1} - A_X)^{-1} \delta_x, \delta_y }, \nonumber \\ \label{eq:ineq of generator}
\end{eqnarray}
for any $x, y \in VX$, where $C_2$ is a positive constant satisfying
\begin{equation}
R e^{\beta R} \int_\bbR \abs{ p(t) } e^{- \frac{(t-a)^2}{b}}  e^{t R} dt < C_2.
\end{equation}
By the above inequality (\ref{eq:ineq of generator}), we get
\begin{eqnarray}
& & \sum_{y \in V \Lambda_n} \abs{ \innpro{e^{\beta H} \int_\bbR p(t) e^{- \frac{(t-a)^2}{b}} e^{itH} \delta_x dt, \delta_y} } v(y) \leq C_2 \sum_{y \in V\Lambda_n} \innpro{ (R \mathbbm{1} - A_X)^{-1} \delta_x, \delta_y } v(y) \nonumber\\
&=& C_2 \innpro{ (R\mathbbm{1} - A_X)^{-1} \delta_x, v \restriction_{V\Lambda_n} } = C_2 \sum_{k = 0}^\infty \frac{\innpro{A_X^k \delta_x, v\restriction_{V\Lambda_n}}}{R^{k+1}} \nonumber\\
&\leq& C_2 (R - \norm{A_X})^{-1} v(x). \label{eq:well-definedness of paring}
\end{eqnarray}
Finally, we show the latter part of the condition $3^\prime$.
For any $f \in \frah_2$, by definition of $v$,
\begin{equation}
\innpro{ e^{\beta H} f, v } = \innpro{f,v}.
\end{equation}
Thus, we are done. 
\QED

\begin{theorem} \label{theorem:factor decomp h}
Suppose that the adjacency operator $A_X$ of a graph $X$ is transient.
For $D > 0$, the two-point function $\varphi_D$ defined in {\rm (\ref{eq:BECoccur})} is a non-factor KMS state on $\calW(\frah_1, \sigma)$ or $\calW(\frah_2, \sigma)$.
Moreover, we have factor decomposition of $\varphi_D$ into extremal KMS states
\begin{equation}
\varphi_D = \frac{1}{2 \pi} \int_{\bbR^2} \varphi_{s_1, s_2} e^{- \frac{s_1^2 + s_2^2}{2}} ds_1 ds_2. \label{eq:factor decomp h}
\end{equation}
\end{theorem}

\noindent {\bf Proof.}
Since $\innpro{ \cdot, (e^{\beta H} + \mathbbm{1})(e^{\beta H} - \mathbbm{1})^{-1} \cdot }$ is positive definite inner product on $\frah_1$ and $\frah_2$, it suffice to show that $\innpro{v,f}$, $f \in \frah_1$ or $f \in \frah_2$ is not continuous with respect to the norm $\innpro{ \cdot, (e^{\beta H} + \mathbbm{1})(e^{\beta H} - \mathbbm{1})^{-1} \cdot }$ by Theorem \ref{theorem:disjointness} and \ref{theorem:factor decomposition}.
Let $p_n$ be the polynomial defined by
\begin{equation}
p_n(x) = \sum_{k = 0}^n \frac{(- n x)^k}{k!}.
\end{equation}
For any $f \in \frah_1$, $(p_n(H) - \mathbbm{1}) f \in \frah_1$.
Put $f_n = (p_n(H) - \mathbbm{1}) f$.
Then 
\begin{equation}
\innpro{ f_n - f, (e^{\beta H} + \mathbbm{1})(e^{\beta H} - \mathbbm{1})^{-1} (f_n - f) } \to 0, \quad (n \to \infty)
\end{equation}
and
\begin{equation}
\innpro{v, f_n} = 0
\end{equation}
for any $n \in \bbN$.
Thus, we have that $\innpro{v, \cdot}$ is not continuous.

For any $f \in \frah_2$, we put $f_n = p_n(H) f$.
We can show $f_n \in \frah_2$.
Similarly the case of $\frah_1$, we can prove the statement.
\QED

\section*{Acknowledgments}
The author would like to thank Professor Taku Matsui for discussions and comments for this work.


\begin{thebibliography}{0}

\bibitem{Araki1} H. Araki, M. Shiraishi,
On quasifree states of the canonical commutation relations. I,
{\it Publ. Res. Inst. Math. Sci.} {\bf 7} (1971/72), 105--120.



\bibitem{Araki2} H. Araki,
On quasifree states of the canonical commutation relations. II,
{\it Publ. Res. Inst. Math. Sci.} {\bf 7} (1971/72), 121--152.


\bibitem{ArakiYamagami} H. Araki, S. Yamagami,
On quasi-equivalence of quasifree states of the canonical commutation relations.
{\it Publ. Res. Inst. Math. Sci.} {\bf 18} (1982), no. 2, 703--758





\bibitem{BratteliRobinsonI} O. Bratteli, D. Robinson,
Operator algebras and quantum statistical mechanics I, 2nd edition
(Springer, 1986.)

\bibitem{BratteliRobinsonII} O. Bratteli, D. Robinson,
Operator algebras and quantum statistical mechanics II, 2nd edition
(Springer, 1997.)




\bibitem{Fidaleo11} F. Fidaleo, D. Guido, T. Isola,
Bose-Einstein condensation on inhomogeneous amenable graphs.
{\it Infin. Dimens. Anal. Quantum Probab. Relat. Top.} {\bf 14} (2011), no. 2, 149--197.


\bibitem{Fidaleo12} F. Fidaleo,
Harmonic analysis on Cayley trees II: The Bose-Einstein condensation.
{\it Infin. Dimens. Anal. Quantum Probab. Relat. Top.} {\bf 15} (2012), no. 4, 1250024, 32 pp. 


\bibitem{Fidaleo15} F. Fidaleo, 
Harmonic analysis on inhomogeneous amenable networks and the Bose-Einstein condensation.
{\it J. Stat. Phys.} {\bf 160} (2015), no. 3, 715--759. 


\bibitem{Honegger90} R. Honegger,
Decomposition of positive sesquilinear forms and the central decomposition of gauge-invariant quasi-free states on the Weyl-algebra.
{\it Z. Naturforsch.} {\bf A 45} (1990), no. 1, 17--28.

\bibitem{HoneggerArmin} R. Honegger, A. Rapp,
General Glauber coherent states on the Weyl algebra and their phase integrals.
{\it Phys. A} {\bf 167} (1990), no. 3, 945--961.


\bibitem{LewisPule}
J. T. Lewis, J. V. Pul\`{e},
The equilibrium states of the free Boson gas. 
{\it Comm. Math. Phys.} {\bf 36} (1974), 1--18. 


\bibitem{ManuceauVerbeure68} J. Manuceau, A. Verbeure,
Quasi-free states of the C.C.R.-algebra and Bogoliubov transformations.
{\it Comm. Math. Phys.} {\bf 9} (1968), 293--302.

\bibitem{ManuceauRoccaTestard69}
J. Manuceau, F. Rocca, D. Testard,
On the product form of quasi-free states.
{\it Comm. Math. Phys.} {\bf 12} (1969), 43--57.


\bibitem{Matsui} T. Matsui,
BEC of free bosons on networks.
{\it Infin. Dimens. Anal. Quantum Probab. Relat. Top.} {\bf 9} (2006), no. 1, 1--26. 


\bibitem{PuleVerbeureZagrebnov}
J. V. Pul\'{e}, A. Verbeure, V. A. Zagrebnov,
On nonhomogeneous Bose condensation.
{\it J. Math. Phys.} {\bf 46} (2005), no. 8, 083301, 8 pp.


\bibitem{RoccaSirugueTestard}
F. Rocca, M. Sirugue, D. Testard,
On a class of equilibrium states under the Kubo-Martin-Schwinger condition. II. Bosons.
{\it Comm. Math. Phys.} {\bf 19} (1970), 119--141.

\bibitem{Simon}
B. Simon,
Functional integration and quantum physics. Second edition. 
AMS Chelsea Publishing, Providence, RI, 2005. xiv+306 pp. ISBN: 0-8218-3582-3

\bibitem{vanDaele}
A. van Daele,
Quasi-equivalence of quasi-free states on the Weyl algebra.
{\it Comm. Math. Phys.} {\bf 21}, (1971), 171--191.

\bibitem{Yamagami10} S. Yamagami,
Geometry of quasi-free states of CCR algebras. {\it Internat. J. Math.} {\bf 21} (2010), no. 7, 875--913.


\bibitem{Yamagami12} S. Yamagami, 
Geometry of coherent states of CCR algebras.
{\it Infin. Dimens. Anal. Quantum Probab. Relat. Top.} {\bf 15} (2012), no. 2, 1250009, 9 pp.



\end{thebibliography}
\end{document}